\documentclass[pra,aps,10pt,amsmath,amssymb,twocolumn,superscriptaddress]{revtex4-1}

\usepackage{graphicx}
\usepackage{dcolumn}
\usepackage[dvipsnames,table,xcdraw]{xcolor}
\usepackage{bm}
\usepackage[colorlinks=true]{hyperref} 
\usepackage[export]{adjustbox}
\usepackage{braket}

\begin{document}
\preprint{APS/123-QED}

\title{Synthetic-lattice Bloch wave dynamics in a single-mode microwave resonator}

\author{F. Ahrens}
\email{fahrens@fbk.eu}
\thanks{these authors contributed equally to this work.}
\affiliation{Fondazione Bruno Kessler (FBK), I-38123, Trento, Italy}

\author{N. Crescini}
\email{ncrescini@fbk.eu}
\thanks{these authors contributed equally to this work.}
\affiliation{Fondazione Bruno Kessler (FBK), I-38123, Trento, Italy}

\author{A. Irace}
\affiliation{Fondazione Bruno Kessler (FBK), I-38123, Trento, Italy}
\affiliation{Dipartimento di Fisica, Università di Milano-Bicocca, I-20126, Milano, Italy}

\author{G. Rastelli}
\email{gianluca.rastelli@ino.cnr.it}
\affiliation{Pitaevskii BEC Center, CNR-INO and Dipartimento di Fisica, Universit\`a di Trento, I-38123 Trento, Italy}

\author{P. Falferi}
\affiliation{CNR-IFN, I-38123, Trento, Italy}
\affiliation{Fondazione Bruno Kessler (FBK), I-38123, Trento, Italy}

\author{A. Giachero}
\affiliation{Dipartimento di Fisica, Università di Milano-Bicocca, I-20126, Milano, Italy}
\affiliation{INFN - Milano Bicocca, I-20126 Milano, Italy}
\affiliation{Bicocca Quantum Technologies (BiQuTe) Centre, I-20126 Milano, Italy}

\author{B. Margesin}
\affiliation{Fondazione Bruno Kessler (FBK), I-38123, Trento, Italy}

\author{R. Mezzena}
\affiliation{Dipartimento di Fisica, Università di Trento, I-38123, Trento, Italy}

\author{A. Vinante}
\affiliation{CNR-IFN, I-38123, Trento, Italy}
\affiliation{Fondazione Bruno Kessler (FBK), I-38123, Trento, Italy}

\author{I. Carusotto}
\email{iacopo.carusotto@ino.cnr.it}
\affiliation{Pitaevskii BEC Center, CNR-INO and Dipartimento di Fisica, Universit\`a di Trento, I-38123 Trento, Italy}

\author{F. Mantegazzini}
\email{fmantegazzini@fbk.eu}
\affiliation{Fondazione Bruno Kessler (FBK), I-38123, Trento, Italy}

\begin{abstract}
Frequency-based synthetic dimensions are a promising avenue for expanding the dimensionality of photonic systems. 
In this work, we show how a tilted synthetic lattice is naturally realized by periodically modulating a single-mode resonator under a coherent monochromatic drive. We theoretically study the Bloch wave dynamics in the tilted synthetic lattice, which gives rise to peculiar features in the spectral distribution of the cavity field.
Our predictions are experimentally confirmed using a tunable superconducting microwave resonator. 
\end{abstract}
                           
\maketitle 

The concept of synthetic dimensions~\cite{hazzard2023synthetic} is attracting ever-growing attention in both the atomic quantum gas~\cite{boada:prl2012} and the photonics~\cite{ehrhardt2023perspective} communities as an efficient way to increase the effective dimensionality of a physical system and realize unprecedented topological models~\cite{RevModPhys.91.015006} and states of matter~\cite{ozawa2019topological}, as well as to develop novel schemes of coherent wave manipulation~\cite{Pozar:882338}. In the atomic context, the extra dimension is typically encoded in either the internal~\cite{mancini2015observation,stuhl2015visualizing} or the momentum degrees of freedom~\cite{an2018correlated} and up to four-dimensional geometries have started to be theoretically and experimentally investigated~\cite{price2015four,bouhiron2024realization}. In the photonic context, a variety of one- or two-dimensional, Hermitian or non-Hermitian lattices have been realized in the frequency domain using the series of evenly spaced modes of a ring or a fiber-loop cavity, the inter-mode hopping being typically induced via a temporal modulation of the cavity properties~\cite{ozawa2016synthetic,yuan2018synthetic,lustig2019photonic,wang2021generating,cheng2023multi,pellerin2024wave,peyruchat2024landauzenerqubitunveilingmultiphoton}.

In this work, we apply the concept of synthetic dimensions to the dynamics of a multi-frequency electromagnetic field in a simple system based on a periodically modulated single-mode resonator. Here, the frequency-domain synthetic lattice is encoded in the sidebands of the monochromatic incident field: in contrast to physical lattices, the synthetic nature allows the inter-site hopping to have a long-range character that can be controlled via the waveform of the periodic modulation~\cite{cheng2023multi,pellerin2024wave}. As a novel feature, the different detuning of each frequency component from the single cavity mode naturally introduces a linear potential gradient along the lattice and opens the way to studying phenomena inspired by electron transport in solid-state physics~\cite{ashcroft1976solid}. In particular, signatures of Bloch oscillations \cite{leo1992observation,dahan1996bloch,sapienza2003optical} are found in the spectral distribution of the cavity field under a monochromatic drive, such as the appearance of sidebands at a detuning that is controlled in a continuous way by the amplitude of the modulation up to values much larger than the cavity linewidth. 

As a specific example of implementation, our theoretical predictions are experimentally confirmed in a tunable single-mode superconducting microwave resonator~\cite{Sandberg_APL_2008,yamamoto_flux-driven_2008,zakka-bajjani_quantum_2011,wilson_observation_2011,Silveri_2015,Silveri_2017,Shumeiko2013}, whose resonance frequency is tuned via the magnetic-flux-dependence of a Superconducting Quantum Interference Device (SQUID) embedded in the resonator~\cite{yamamoto_flux-driven_2008}. Complementing the established photon energy lifting scheme~\cite{Gaburro_Optica_2006,preble_changing_2007,Sandberg_APL_2008}, our experimental observations show the potential of the synthetic lattice framework as a way to design new schemes for the spectral manipulation of electromagnetic signals.

The structure of the article is as follows. In Sec.\,\ref{sec:theory} we develop the general theory and introduce the frequency-space synthetic dimension model. Numerical simulations for the dynamics in the tilted synthetic lattice are presented in Sec.\,\ref{sec:numerical}. The experimental setup is introduced in Sec.\,\ref{sec:expt_setup} and the experimental data are summarized in Sec.\,\ref{sec:expt_results}. Finally, conclusions are drawn in Sec.\,\ref{sec:conclu}. Additional material is presented in the Appendices: Additional information on the device and the experimental setup is given in App.\,\ref{app:expt}. Theoretical predictions and experimental observations for sinusoidal modulation are given in App.\,\ref{app:sinu}. Additional theoretical considerations are finally summarized in App.\,\ref{app:theory}.

\section{Theoretical model}
\label{sec:theory}

We consider a coherently driven, single-mode oscillator whose resonance frequency can be externally controlled by temporally varying an external parameter. 

Within a rotating wave approximation (RWA), we can model the cavity field dynamics in terms of a linear equation of motion for the field amplitude $\alpha(t)$
\begin{equation}
  i \dot{\alpha} \! \left( t \right)
  = \left[\omega_0+\delta\omega(t)- i \frac{ \gamma }{ 2 }\right] \alpha \! \left( t \right)
  + E_\mathrm{in}(t) \, ,
  \label{eq:dynamics}
\end{equation}
where $\gamma$ is the loss rate, $\omega_0$ is the bare cavity frequency, and $\delta\omega(t)$ is the externally-determined, time-dependent modulation of the cavity frequency. Here, the RWA is accurate, provided that the modulation occurs on a very slow timescale compared to $\omega_0$, so that no additional photons can be parametrically generated by the modulation~\cite{wilson_observation_2011,lahteenmaki2013dynamical}.

In what follows, we assume that the external incident field driving the cavity is monochromatic, $E_\mathrm{in}(t)=E_\mathrm{in}\,e^{-i\omega_\mathrm{in} t}$, and the modulation $\delta\omega(t)$ is periodic with period $T$ and frequency $\Omega=2\pi/T$ with an arbitrary waveform. 

Via frequency mixing processes, such a modulation creates a series of sidebands around the incident frequency $\omega_\mathrm{in}$, spaced by $\Omega$. Based on this physical insight, we look for a solution for the steady state of Eq.\,\eqref{eq:dynamics} in terms of the multi-frequency ansatz
\begin{equation}
    \alpha\left(t \right) = \sum_n \alpha_n e^{-i(\omega_\mathrm{in}+ n \Omega) t}\,,
    \label{eq:ansatz}  
\end{equation}
where $\alpha_n$ is the complex-valued amplitude of the frequency component at $\omega_\mathrm{in}+n\Omega$ with $n\in\mathbb{Z}$. These spectral components are visible in the spectrum of the steady-state radiation emitted by the cavity. 

\begin{figure}
\centering
\includegraphics[width=\columnwidth]{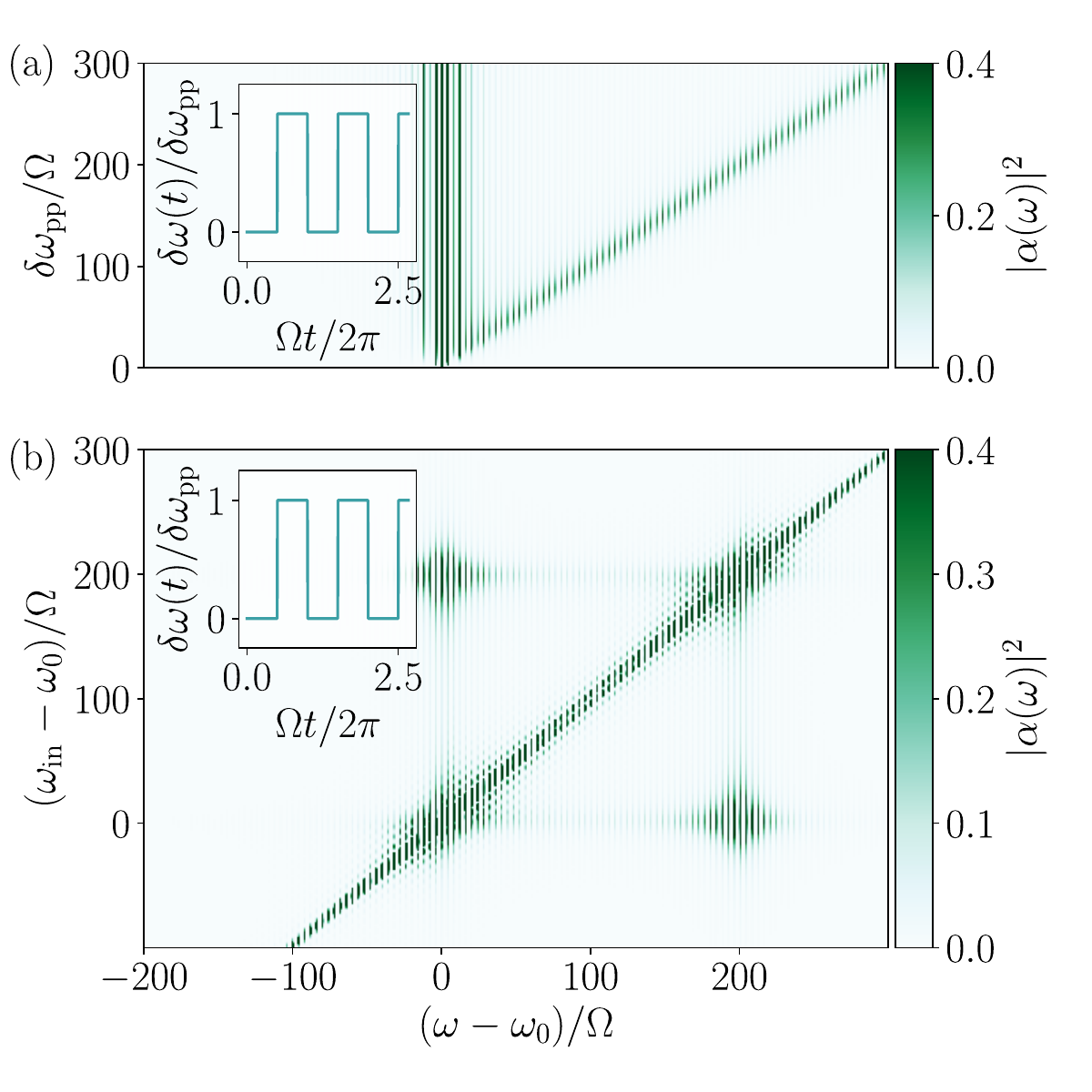}
\caption{Intensity $|\alpha_n|^2$ of the different spectral components of the cavity emission at $\omega=\omega_0+n\Omega$ for the case of the square-wave modulation shown in the inset. Panel (a) illustrates the dependence on the modulation intensity $\delta\omega_{\rm pp}$ for a fixed drive frequency on resonance with the bare cavity $\omega_\mathrm{in}=\omega_0$. Panel (b) shows the dependence on the drive frequency $\omega_{\rm in}$ for a fixed modulation intensity $\delta\omega_\mathrm{pp}$. The linewidth is $\gamma/\Omega=2$ in both panels. In (a) $\omega_\mathrm{in}=\omega_0$, while in (b) $\delta\omega_{\rm pp}/\Omega=200$. 
}
\label{fig:simulations}
\end{figure}

By inserting this ansatz into the motion equation \eqref{eq:dynamics}, we generate an infinite hierarchy of linear equations coupling the amplitudes $\alpha_n$,
\begin{equation}
\left[
 \omega_\mathrm{in} - \omega_0 + n\Omega+i\frac{\gamma}{2}\right]\,\alpha_n - \sum_{m\neq 0} \delta\omega_m \alpha_{n-m} = E_\mathrm{in}\delta_{n,0}\,
 \label{eq:hierarchy}
\end{equation}
where $\delta\omega_m$ are the Fourier components of the slow modulation $\delta\omega(t)=\sum_m \delta\omega_m\,e^{-im\Omega t}$.   
\begin{figure}[hbtp]
\centering
\includegraphics[width=\columnwidth]{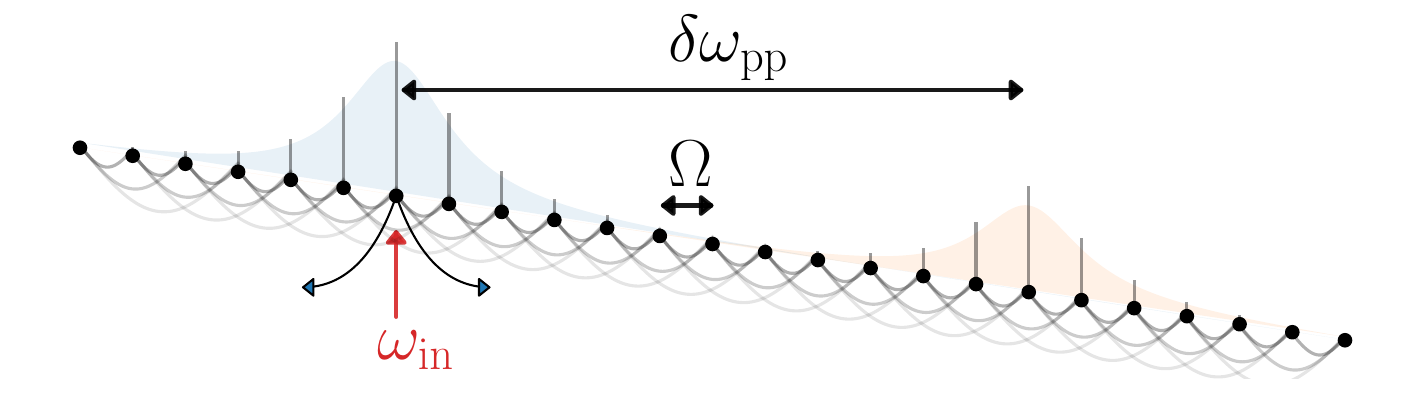}
\caption{Sketch of the tilted frequency-space synthetic lattice encoded by Eq.\,(\ref{eq:hierarchy}), with $\omega_\mathrm{in}=\omega_0$. The light blue and light red shaded areas indicate the central stripe and the side-stripe that appear under a square-wave modulation as shown in Fig.\,\ref{fig:simulations}. 
}
\label{fig:tiltedlattice}
\end{figure}
This equation can be mapped into the stationary solution of a (driven-dissipative) wave mechanics problem of a particle in a one-dimensional lattice displaying long-range hopping of amplitude $\delta\omega_m$ for a $m$-site jump and a constant potential gradient of strength $\Omega$, as summarized in the Wannier-Stark Hamiltonian
\begin{equation}   
\tilde{H}_\mathrm{WS}
\!
=  
\!
\sum_n 
\!
\Big[-\Omega n \ket{n}\bra{n} 
\!
+
\!
\sum_{m}
\!
\delta\omega_m \ket{n}\bra{n+m}
\Big]
\label{eq:WS}
\end{equation}
and in the sketch in Fig.\,\ref{fig:tiltedlattice}.
In the absence of the external potential, the Bloch band dispersion resulting from the hopping term would display a shape that, as a function of the ``crystal'' momentum $k\in[-\pi,\pi]$, exactly recovers the temporal profile of the modulation,
$\mathcal{E}(k)=\sum_m \delta\omega_m\,e^{-im k}=\delta\omega(k T/2\pi)$~\cite{Piccioli:PRA2022}. An example of such a dispersion is shown in the insets of Fig.\,\ref{fig:simulations}.

An analytical study of the model based on Eq.\,\eqref{eq:hierarchy} based on the propagator of Eq.\,\eqref{eq:WS} is given in App.\,\ref{app:theory} but does not provide much insight into the physical mechanisms. This can rather be obtained with the following intuitive arguments based on a semiclassical theory of electron transport in solids~\cite{ashcroft1976solid}.
Neglecting for the moment the driving $E_{\rm in}$ and the losses $\gamma$, the particle moves at each time with a group velocity $d\mathcal{E}(k)/dk$ evaluated at the instantaneous value of the wave vector $k(t)$. As $k(t)$ evolves along the band dispersion at a rate $dk/dt=\Omega$ under the effect of the uniform force, the motion will display periodic Bloch oscillations~\cite{dahan1996bloch,sapienza2003optical,leo1992observation}
\begin{equation}
\Delta \bar{n}(t)  
=\int^{t}_0 dt'\, \frac{d\mathcal{E}(k(t'))}{d k}
=\frac{1}{\Omega}
\left[\mathcal{E}\left(k(t)\right) - \mathcal{E}\left(k(0)\right) \right]
\,,
\label{eq:semicl}
\end{equation}
where $\Delta\bar{n}$ is measured from the initial location and $k(0)$ is the initial wave vector of the particle.

As a peculiar feature of our synthetic dimension scheme, hopping can include arbitrarily long-range terms so that arbitrary band dispersions can be engineered. As a most interesting example, we consider an extreme scenario with a square-wave modulation with $\delta\omega(t)=\delta\omega_\mathrm{pp}\,\Theta\left[\cos\left(\Omega t\right)\right]$, where the Fourier components display a long tail $\delta\omega_{m} = \delta\omega_\mathrm{pp} \sin(\pi m/2)/(\pi m)$ and $\delta\omega_{m=0}=\delta\omega_\mathrm{pp}/2$. 
An analogous study for the (less intriguing) case of a sinusoidal modulation is given in App.\,\ref{app:sinu}.

In any realistic experimental implementation of a square wave, the sharp edges are smoothened out by the finite bandwidth of the modulation process: This guarantees that no parametric emission process takes place, while the rise time is still short enough (i.e.~on the scale of $1/\gamma$) to preserve the peculiar Bloch dynamics. 
Inserting the square wave into Eq.\,\eqref{eq:semicl} and
assuming the initial wave vector $k(0)=0$ on the lower part of the band dispersion, the Bloch oscillations
\begin{equation}
\Delta \bar{n}(t)  
=  
\frac{\delta\omega_\mathrm{pp}}{\Omega} \,
 \Theta\left[ k(t) - \frac{\pi}{2}\right]\,,
\label{eq:BO}
\end{equation}
display a series of sudden upward jumps at the discontinuity points of the band dispersion.

On top of this conservative dynamics, we have to consider that according to our driven-dissipative equation \eqref{eq:hierarchy} particles have a finite lifetime $\gamma^{-1}$ and are continuously injected into the central site $n=0$ at an energy $\omega_{\rm in}$. As a result, the injection of particles will be most efficient into all those $k$ states that satisfy the resonance condition $\omega_0+\mathcal{E}(k)=\omega_{\rm in}$. In the square modulation case under examination, setting $\omega_{\rm in}\simeq \omega_0$ concentrates the population on the lower part of the band dispersion around $k(0)=0$. According to Eq.\,\eqref{eq:BO}, the Bloch oscillation process will then displace a part of the population creating a sideband shifted by $\delta\omega_{\rm pp}$.

\section{Numerical simulations}
\label{sec:numerical}

This semiclassical prediction can be validated by numerically solving the set of equations (\ref{eq:hierarchy}), so to obtain the stationary values of the amplitudes $\alpha_n$ of the different frequency sidebands. 
Suitable cut-offs $m_c$ and $n_c$ model the smoothened edges of the square wave. Fig.\,\ref{fig:simulations}a shows the result of such a calculation for the case of a square-wave modulation considered above under the resonance condition $\omega_\mathrm{in}=\omega_0$.

As expected from the semiclassical theory, the color-plot as a function of the modulation amplitude $\delta\omega_{\rm pp}$ displays a main signal around the incident frequency $\omega_{\rm in}=\omega_0$, as well as an additional, well-isolated side-stripe at a detuning that grows proportionally to the modulation intensity $\delta\omega_\mathrm{pp}$. Most remarkably, the frequency shift of this sideband can reach arbitrarily large values well beyond the cavity linewidth and the frequency of the modulation. In contrast to typical wave-mixing processes, the amplitude of the frequency shift is controlled here by the strength of the modulation signal rather than by its frequency.

Even richer features are visible in the color-plot shown in Fig.\,\ref{fig:simulations}b displaying the $\omega$-dependent emission spectrum (horizontal axis) for different values of the incident frequency $\omega_\mathrm{in}-\omega_0$ (vertical axis): the diagonal stripe corresponds to an emission at the incident frequency $\omega_\mathrm{in}$, while the horizontal and vertical stripes correspond to the bare cavity frequency $\omega_0$ and the shifted cavity frequency $\omega_0+\delta\omega_\mathrm{pp}$, respectively. 
In addition, this plot can be interpreted in terms of the semiclassical theory presented above. In fact, the injection is most efficient when $\omega_{\rm in}$ is resonant with the lower or upper part of the dispersion, respectively, at $\omega_{\rm in}=\omega_0$ or $\omega_0+\delta\omega_{\rm pp}$. For each of these values, the other isolated spot that is visible in the figure corresponds to the upward or downward shift caused by the Bloch oscillation process. The diagonal line describes emission at the incident frequency $\omega_{\rm in}$ and can be understood as the result of a (possibly non-resonant) injection of waves into a quasi-static cavity.

From a different perspective, these results for a square-wave modulation can be interpreted as a photon energy lifting phenomenon theoretically discussed in~\cite{Gaburro_Optica_2006} and experimentally pioneered in~\cite{preble_changing_2007,Sandberg_APL_2008}: the vertical stripes originate from the ring-down of the cavity field at its instantaneous natural frequency (i.e., $\omega_0$ or $\omega_0+\delta\omega_\mathrm{pp}$) after its frequency has been suddenly changed (downward or upward).

\begin{figure}
\includegraphics[width=.48\textwidth]{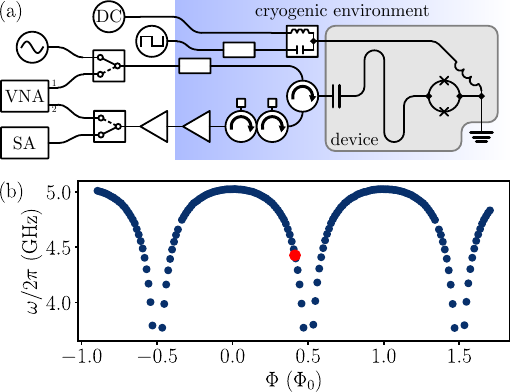}
\caption{Experimental setup. (a) Schematic of the resonator device (gray box) and of the setup used to characterize and operate it. (b) Modulation of the resonance frequency of the tunable resonator as a function of the magnetic flux $\Phi$ in the SQUID. The working point used in the experiments is marked in red.}
\label{fig:setup}
\end{figure}

\begin{figure*}
\includegraphics[width=.9\linewidth]{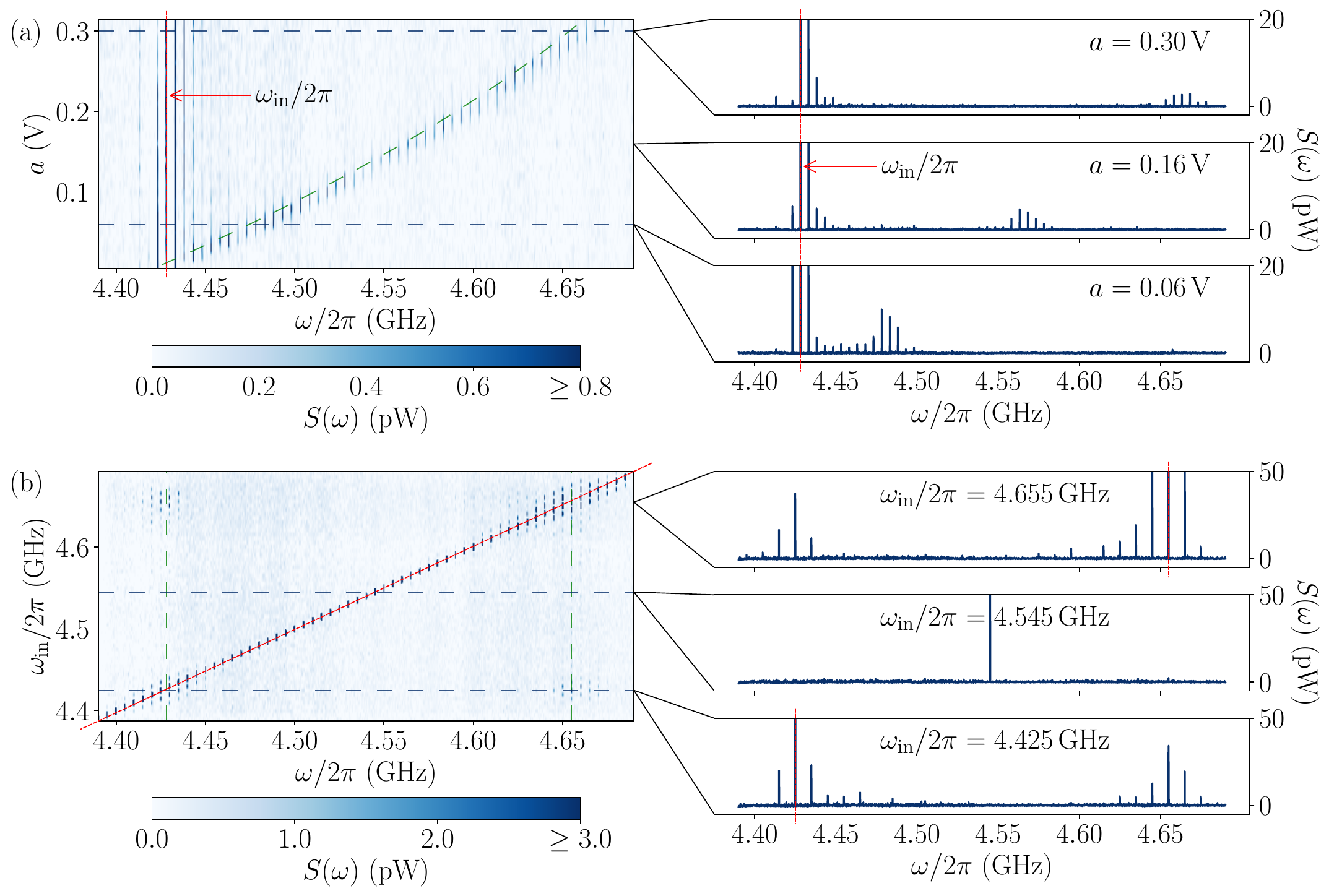}
\caption{Experimental results. (a) Color-plot of the detected power spectrum as a function of the amplitude $a$ of a square-wave modulation at frequency $\Omega=5\,\mathrm{MHz}$. 
The red line indicates the input frequency, while the dashed green line shows the expected frequency of the second maximum of the output power spectrum. The side panels show cuts of the color-plot for three different values of the modulation amplitude $a$. The input frequency is indicated by a red line.
(b) Color-plot of the detected power spectrum as a function of the incident frequency $\omega_\mathrm{in}$ for a square-wave modulation of frequency $\Omega=10\,\mathrm{MHz}$ and an amplitude $a=0.3\,\mathrm{V}$. The red line indicates the input frequency. The side panels show cuts of the color-plot for three different values of the incident frequency $\omega_\mathrm{in}$, as indicated by the red lines.
}
\label{fig:photon_energy_lifting}
\end{figure*}

\section{The experimental setup}
\label{sec:expt_setup}

These theoretical predictions are put into practice using a superconducting circuit in which the single-mode cavity is realized as a planar microwave resonator, whose resonance frequency can be tuned in a controlled way. 
The device features an aluminum coplanar waveguide resonator terminated to ground via a SQUID (Superconducting Quantum Interference Device). 
The SQUID is formed by two Al-AlO$_x$-Al overlap Josephson junctions~\cite{bal_overlap_2021, wu_APL_2017} with an area of $3\times3$\,\textmu m$^2$ and a critical current of about 2\,\textmu A.
By modulating the magnetic flux threading the SQUID loop, the inductance of the resonant circuit is changed, and consequently, its resonance frequency $\omega$ is modulated. The schematics of the experimental setup are shown in Fig.\,\ref{fig:setup}a. Further details on the layout and microfabrication of the device are provided in App.\,\ref{subsec:design_and_fab}.

The device is housed in a dilution refrigerator with a base temperature of about 30\,mK, and is measured in reflection with a low-noise spectroscopy apparatus connected to the device through a circulator. 
The spectroscopy apparatus consists of an attenuated input line and an amplified readout line, as schematically shown in Fig.\,\ref{fig:setup}a.
The flux through the SQUID can be tuned via a modulation line coupled to it. The modulation line includes a filtered DC branch and an attenuated RF branch using a bias-tee, to separately control the DC and microwave flux.
The experiments are performed with a vector-network analyzer (VNA) and a spectrum analyzer (SA), for characterization and power measurements, respectively. 
The detailed experimental apparatus is described in App.\,\ref{subsec:apparatus_detailed}.

The bare frequency of the resonator is $5.0\,\mathrm{GHz}$, and its quality factor is $Q\simeq 10^3$.
The frequency tunability of the resonator is verified by changing the DC flux $\Phi$ and measuring the corresponding resonance frequency shift, obtaining the expected periodic dependence and a span above 1\,GHz, as shown in Fig.\,\ref{fig:setup}b. 
To optimize the frequency tunability and avoid excessive non-linearity, a working point corresponding to a base frequency $\omega_0=(2\pi)4.43\,\mathrm{GHz}$ is chosen, as indicated in Fig.\,\ref{fig:setup}b.
The modulation tone is supplied by a continuous square-wave generator through the microwave branch of the modulation line which has no DC component (Fig.\,\ref{fig:setup}a). 
The square wave features a rise time of about 20\,ns and a variable amplitude $a$. 
To form a non-negative square-wave modulation at the device level, we apply an active correction of the DC flux for each value of $a$, thereby keeping the base frequency approximately fixed, as detailed in App.\,\ref{subsec:device_chara}.

\section{Experimental results}
\label{sec:expt_results}

Using this experimental setup, we first carry out a series of measurements under a continuous-wave monochromatic excitation resonant with the bare cavity frequency $\omega_0$ in the presence of a square-wave modulation. Additional comparative studies for a sinusoidal modulation are reported in App.\,\ref{app:sinu}.

A color-plot of the resulting output power spectra for different values of the modulation amplitude is reported in Fig.\,\ref{fig:photon_energy_lifting}a. In agreement with the theoretical prediction of Fig.\,\ref{fig:simulations}a, the plot displays a vertical stripe around the driving frequency, as well as a pronounced side-stripe whose frequency shifts almost proportionally to the modulation strength. 
A closer view on this physics is obtained by looking at the cuts of the color-plot. Here, one can clearly observe the main signal around the driving frequency $\omega_\mathrm{in}=\omega_0$ and the side peaks generated by the square-wave modulation. 

In a second series of measurements, we vary the frequency $\omega_\mathrm{in}$ of the monochromatic drive while keeping a constant value $a=0.3\,\mathrm{V}$ for the amplitude of the square-wave modulation, which corresponds to a shift of the cavity resonance around $\delta\omega_\mathrm{pp}\sim 200\,\mathrm{MHz}$.
The results are displayed in Fig.\,\ref{fig:photon_energy_lifting}b and show an overall agreement with the theory of Fig.\,\ref{fig:simulations}b. The output is peaked on the $\omega\simeq \omega_\mathrm{in}$ line, with a maximum intensity when the drive frequency is in resonance with either the bare or the shifted cavity position, indicated by the vertical green dashed lines. When the input frequency $\omega_\mathrm{in}$ is close to either of the two quasi-static cavity frequencies, one can also see evidence of the theoretically predicted emission on the side-stripes. The small discrepancies between the experimental data and the numerical simulations are caused by noise, a non-ideal square wave, and a non-flat baseline.

\begin{figure}[htbp]
\centering
\includegraphics[width=0.48\textwidth]{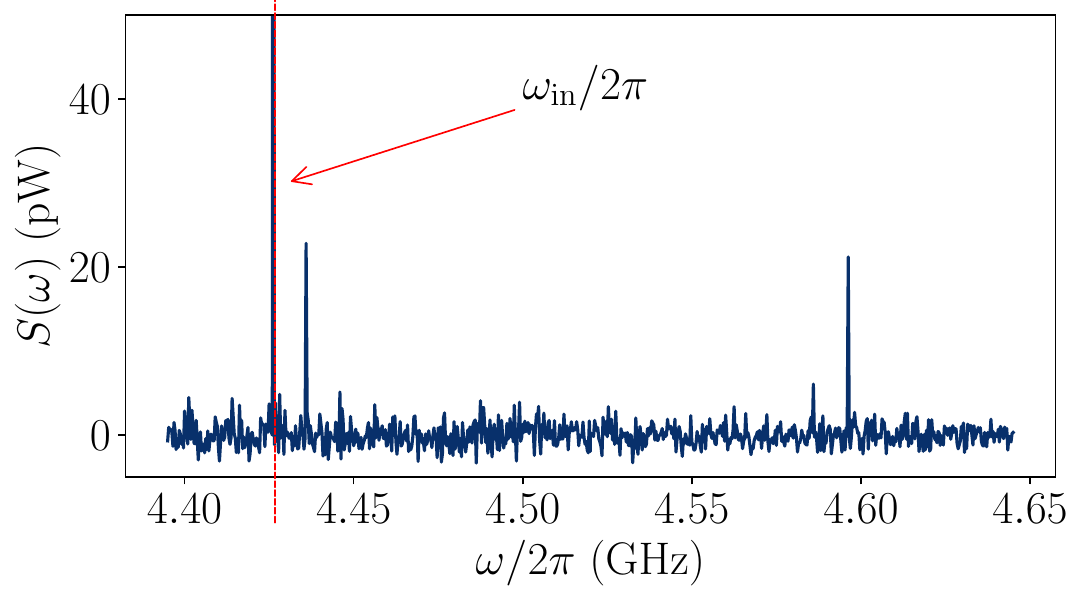}
\caption{Detected power spectrum for a weak incident field in the single photon regime.}
\label{fig:single_photon}
\end{figure}

As a final point, we verify that our predictions are not limited to strong classical fields containing a macroscopic number of photons, but extend directly down to the quantum regime of weak drive fields containing only a very small number of photons. To this end, we reduce the signal power to $\sim-135\,$dBm, so that the average occupation number of the resonator is $\langle N \rangle \simeq 1$. Under this condition, choosing $\Omega = 5\,\mathrm{MHz}$ and $a=0.2\,\mathrm{V}$, with an extended averaging time, we record the spectrum of Fig.\,\ref{fig:single_photon}, which clearly demonstrates that the same spectral features are also visible in a quantum regime.

\section{Conclusions and outlook} 
\label{sec:conclu}
In this work we have demonstrated how a tilted synthetic lattice in the frequency domain naturally arises in the field dynamics in a single-mode resonator periodically modulated in time with a large amplitude and an arbitrary waveform. The flexibility of the elementary frequency-mixing processes induced by the modulation offers a great tunability in the strength and the short- vs.~long-range character of the inter-site hopping in the synthetic lattice.

Building on these elementary elements, a rich dynamics is expected to be observable in the synthetic frequency-domain lattice. For instance, consequences of Bloch oscillations in the tilted lattice are highlighted in the output power spectrum, such as the appearance of sidebands with a frequency detuning much larger than the bare-cavity linewidth. In spite of the elementary nature of the basic frequency-conversion process, the strongly non-perturbative nature of the ensuing dynamics is signaled, for instance, by the fact that the frequency position of the sidebands is not controlled by the modulation frequency, as normally happens in standard frequency-conversion processes, but rather by the modulation amplitude.

Our theoretical predictions are experimentally implemented in a planar tunable superconducting microwave resonator and are verified down to a quantum regime of low photon occupation.
In addition to fundamental studies of wave dynamics using synthetic lattices, our experimental observations open exciting perspectives toward innovative spectral manipulation methods, which could allow for a controlled shape tuning of continuous-wave microwave spectra.

\begin{acknowledgments}
We thank the FBK cleanroom team for support with the microfabrication of the devices. I.C.~is grateful to Tomoki Ozawa and Philippe St-Jean for continuous collaboration on the topic of synthetic dimensions. 
We acknowledge financial support from Q@TN, the joint lab between the University of Trento, FBK-Fondazione Bruno Kessler, INFN-National Institute for Nuclear Physics, and CNR-National Research Council and financed by the Provincia Autonoma di Trento (PAT). The authors also acknowledge the support of the National Quantum Science and Technology Institute through the PNRR MUR Project under Grant PE0000023-NQSTI, co-funded by the European Union - NextGeneration EU. N.C.~and F.M.~are supported by Qu-Pilot SGA (Horizon Europe, Project ID: 10111398) and by MiSS (Horizon Europe, Project ID: 101135868). G.R.~and I.C.~acknowledge financial support from the Provincia Autonoma di Trento (PAT). A.G.~acknowledges support from the Horizon 2020 Marie Skłodowska-Curie actions (H2020-MSCA-IF GA No.101027746).
\end{acknowledgments}


\appendix

\section{Additional experimental material}
\label{app:expt}

In this first Appendix we give more details on the device and on the experimental apparatus used for the measurements.

\begin{figure}[htbp]
\includegraphics[width=0.45\textwidth]{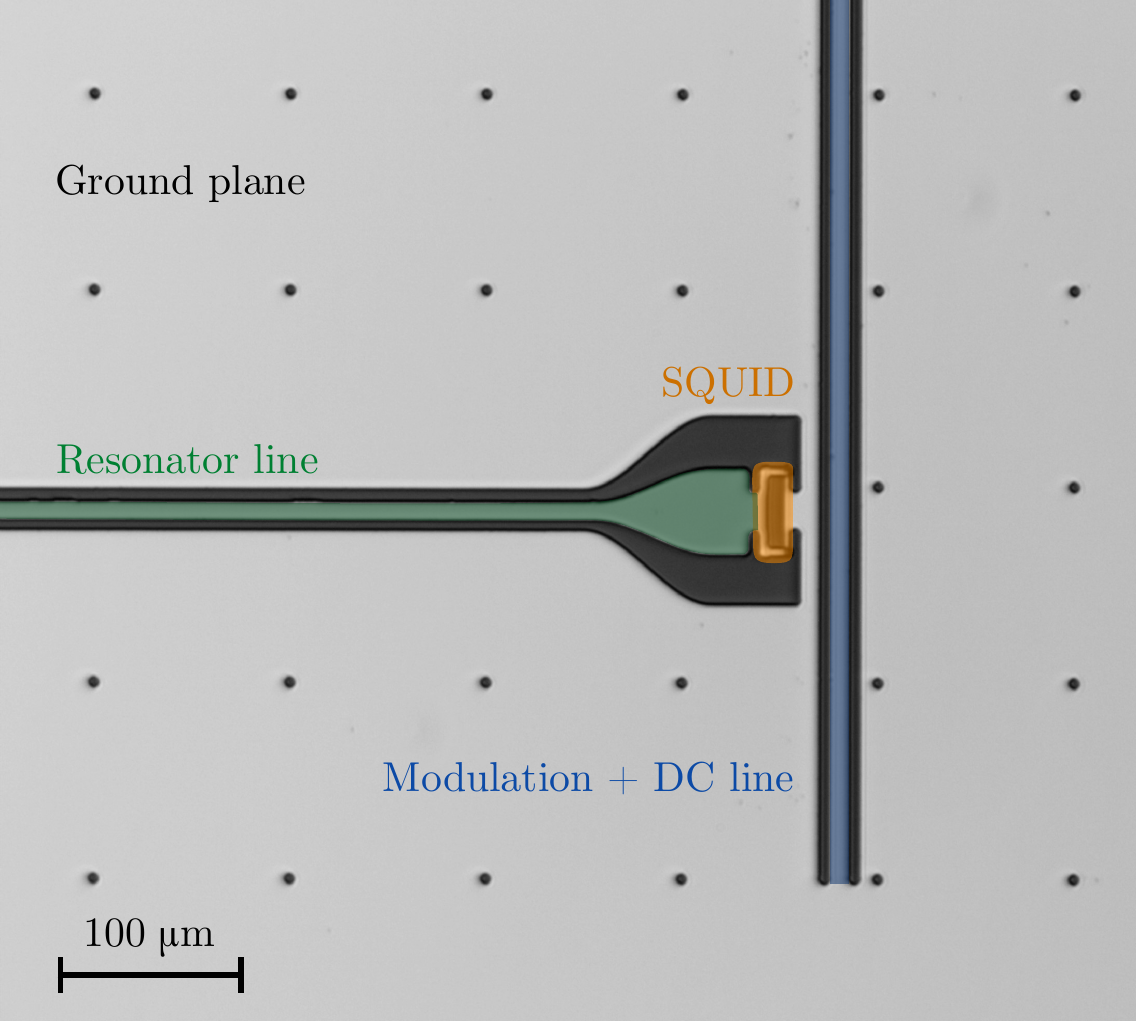}
\caption{Optical microscope photograph of part of the device under investigation. The SQUID-terminated resonator and the modulation and DC line are visible. }
\label{fig:foto_lifter}
\end{figure}

\subsection{Design and microfabrication of the device}
\label{subsec:design_and_fab}

The device comprises an aluminum coplanar waveguide resonator terminated to ground via a SQUID. The SQUID features two Al-AlO$_x$-Al overlap Josephson junctions with an area of $3\times3$\,\textmu m$^2$ and a critical current of approximately 2\,\textmu A. The SQUID loop is inductively coupled to a second waveguide that acts as a flux modulation line. An optical microscope photograph of the SQUID and the flux line is shown in Fig.\,\ref{fig:foto_lifter}. The structures are defined by two UV lithography steps after coating with a positive photoresist, while the aluminum film is deposited via sputtering on a Si substrate. The overlap Josephson junctions are fabricated as follows: (\textit{i}) after Al sputtering, the first layer is defined by lithography and wet aluminum etching, (\textit{ii}) the second lithography then patterns the second layer; (\textit{iii}) the device enters the sputtering chamber again for an argon plasma cleaning and subsequently, with no vacuum breaking, the insulating barrier is formed by injecting oxygen in the chamber, resulting in an oxidation dose of 
$27 \, \mathrm{mbar}\times \mathrm{min}$; (\textit{iv}) the second aluminum layer is then sputtered creating the Josephson junctions, and, as a final step, the structures are defined via lift-off.

\subsection{Detailed experimental apparatus}
\label{subsec:apparatus_detailed}

\begin{figure}
\includegraphics[width=.45\textwidth]{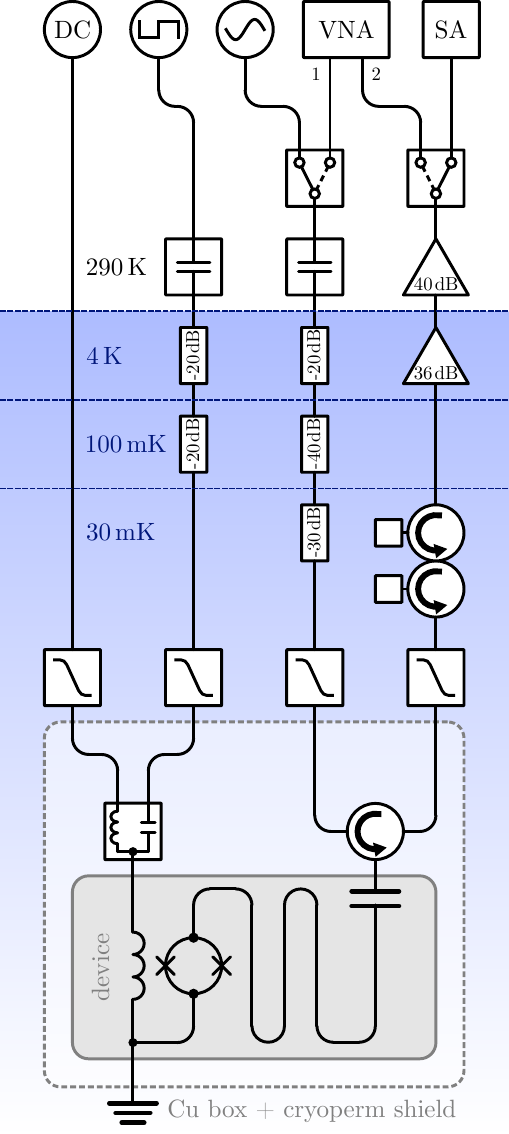}
\caption{Schematics of the room temperature and cryogenic setup of the experiment. The setup is organized to alternatively probe the device under observation through a vector-network analyzer (VNA) or through an external source and a spectrum analyzer (SA).
}
\label{fig:setup_detail}
\end{figure}

The experiment is housed in a dilution refrigerator equipped with a total of three RF lines and one DC line. The setup is organized to alternatively probe the device under observation through a vector-network analyzer (VNA) or through an external monochromatic source and a spectrum analyzer (SA). Detailed schematics of the setup are shown in Fig.\,\ref{fig:setup_detail}. The spectroscopy part of the apparatus, connected to the device via a circulator, consists of an input and a read-out line. The former is gradually attenuated by $-90\,$dB from room temperature to the mixing chamber of the refrigerator, while the latter features two cascaded low-noise transistor amplifiers. The first stage has a nominal noise temperature of 2\,K and is isolated from the sample by a dual-junction isolator. The total gain of the amplification chain is about 76\,dB. 
The modulation line is split in two using a bias-tee, to separately control the DC and microwave flux. The DC line is not attenuated but heavily low-pass filtered with lumped-element and powder filters, while the microwave line is gradually attenuated by $-40\,$dB. All input lines are equipped with infrared filters, and all attenuated microwave lines have DC blocks. The device under investigation and the circulator are hosted inside a copper box which in turn is encased in a Cryoperm box.

\subsection{Additional device characterization}
\label{subsec:device_chara}

\begin{figure}[t]
\includegraphics[width=.48\textwidth]{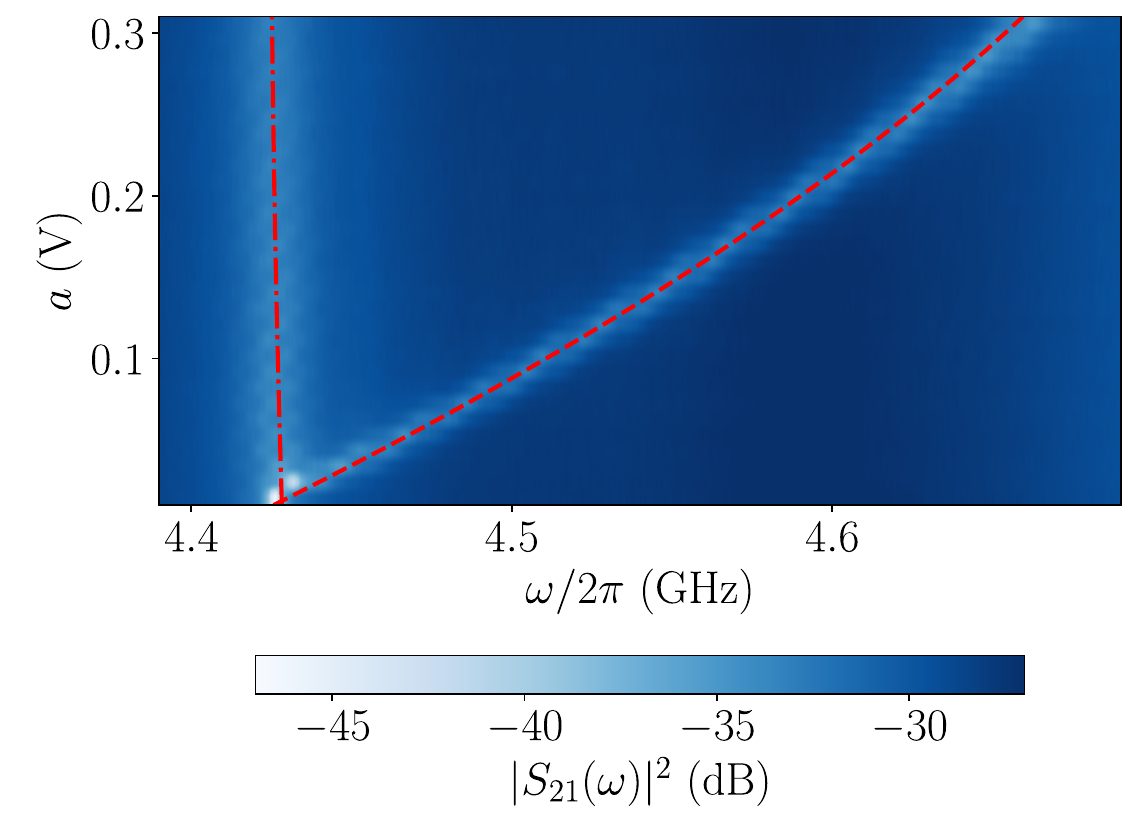}
\caption{Spectroscopy of the modulated system resonances, showing the stability of the base frequency $\omega_0$ and of the shifted frequency $\omega_0+\delta\omega_\mathrm{pp}$ (see text for further details).}
\label{fig:app_s21}
\end{figure}

As mentioned in the main text, to achieve a non-negative square-wave modulation at the device level, we apply an active correction of the DC flux for every value of the square-wave amplitude $a$ to keep the device's base resonance frequency steady within its bandwidth as shown in Fig.\,\ref{fig:photon_energy_lifting}a of the main article. For this, we have performed a calibration, monitoring the reflection of the device as a function of the modulation amplitude $a$ and adjusting the DC bias for each $a$ so that the condition $\omega_0(a)\approx\omega_0(a=0)$ is met for all $a$. In Fig.\,\ref{fig:app_s21} we present the reflection of the device $|S_{21}(\omega)|^2$ as a function of the modulation amplitude $a$ after calibration, measured with a VNA with a resolution bandwidth much smaller than the modulation frequency. We could verify that (\textit{i}) the position of the lifted resonance as a function of the modulation amplitude is as expected (dashed line in Fig.\,\ref{fig:app_s21}), (\textit{ii}) the variation of the base frequency $\omega_0$ is smaller than the resonance bandwidth (dashed-dotted line in Fig.\,\ref{fig:app_s21}), and (\textit{iii}) the quality factors of the resonances do not change significantly with $a$. We remark that Fig.\,\ref{fig:app_s21} is simply a measure of the response function at different frequencies: we scan the frequency input and observe the amplitude of the reflected output at the same frequency, when the modulation tone is on.
The output only reflects the location of the resonances at different times during the modulation and therefore carries no information on the stored photons and their spectral properties.
For the asymmetric square wave used in Fig.\,\ref{fig:app_s21}, the resonance is at $\omega_0$ and $\omega_0 + \delta\omega_\mathrm{pp}$.
As can be seen in the flux modulation plot in Fig.\,\ref{fig:setup}b of the main paper, the frequency behavior is linear only in first approximation. This is the reason why in Fig.\,\ref{fig:app_s21} and in Fig.\,\ref{fig:photon_energy_lifting}a of the main paper the position of the right resonance $\omega_0 + \delta\omega_\mathrm{pp}$ does not follow a straight line, but rather a parabola-like curve. The reason why the base frequency $\omega_0$ appears to slightly drift to the left in Fig.\,\ref{fig:app_s21} is due to an imperfect calibration of the applied DC offset.

\section{Additional results for a sinusoidal modulation}
\label{app:sinu}

In this Appendix, we summarize the results of an analogous study for the case of a sinusoidal modulation, to be compared with the square modulation case reported in the main text.

\begin{figure}[t]
\centering
\includegraphics[width=.48\textwidth]{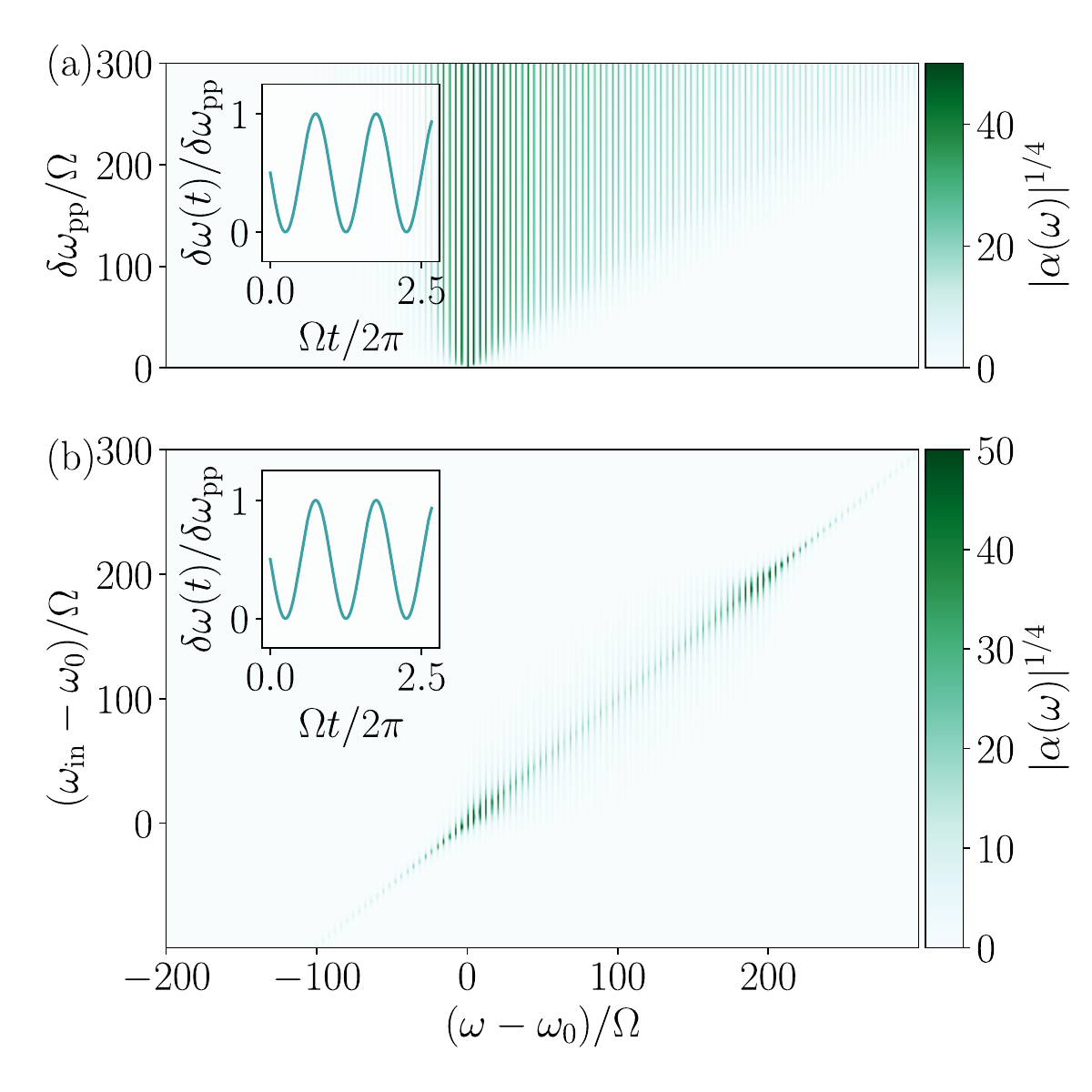}
\caption{Intensity of the spectral components of the emission in the case of the sinusoidal modulation (modulation shown in the insets). Panel (a) illustrates the dependence on the modulation intensity $\delta\omega_{\rm pp}$ for a fixed drive frequency on resonance with the bare cavity $\omega_\mathrm{in}=\omega_0$. Panel (b) shows the dependence on the drive frequency $\omega_{\rm in}$ for a fixed modulation intensity $\delta\omega_\mathrm{pp}$. The linewidth is $\gamma/\Omega= 2$ for all panels. In (a) $\omega_\mathrm{in}=\omega_0$, while in (b) $\delta\omega_\mathrm{pp}/\Omega=200$. Note that we have plotted the fourth root of the field amplitude $|\alpha|$ to highlight the structure of the spectral components.
}
\label{fig:simulations_supp}
\end{figure}

\begin{figure}[t]
\centering
\includegraphics[width=.48\textwidth]{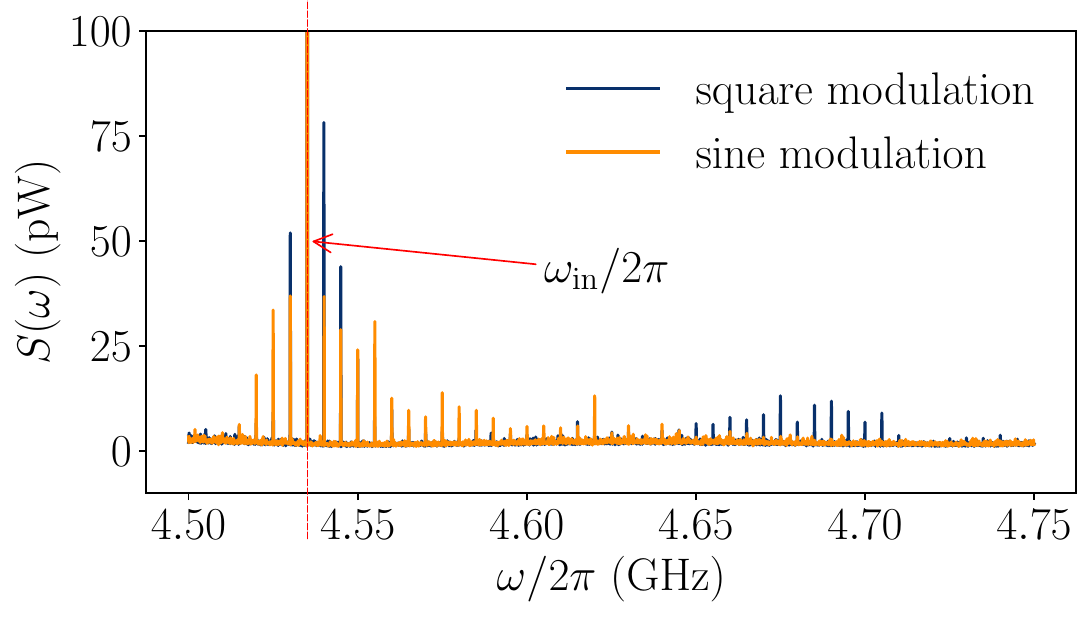}
\caption{Detected power spectrum for the cases of square-wave modulation (dark blue) and sine modulation (dark orange), with an identical peak-peak modulation amplitude and modulation frequency.
}
\label{fig:comparison_sine_square}
\end{figure}

For a sinusoidal modulation of the form $\delta\omega(t)=\delta\omega_\mathrm{pp}\left[\sin\left(\Omega t\right)+1\right]/2$, some resonant states around a few values of $k_0$ are always available for each value of the incident frequency $\omega_\mathrm{in}$ located within the bandwidth of the band dispersion $\mathcal{E}(k)=\delta\omega(k/\Omega)$. From Eq.\,\eqref{eq:semicl}, it is then immediately clear how for different times the possible values of the displacement end up covering all the range of the band dispersion $\mathcal{E}(k)$. This is what happens in the long lifetime limit.

Fig.\,\ref{fig:simulations_supp} shows examples of such spectra for an inverse lifetime $\gamma$ comparable to the Bloch oscillation period. Under a resonant incident field $\omega_\mathrm{in}=\omega_0$ [panel (a)], the spectral distribution remains localized within a wide vertical stripe around $\omega_0$ that, for $\delta\omega_\mathrm{pp}>0$, extends towards higher frequencies. As the finite lifetime limits the displacement that the particles can perform during the Bloch oscillation before being lost, the extension of this asymmetric stripe grows with increasing lifetime $\gamma^{-1}$ until it covers almost uniformly a wide region above $\omega_\mathrm{in}$ up to $\omega_\mathrm{in}+\delta\omega_\mathrm{pp}$ when the lifetime exceeds the Bloch oscillation period.
Varying $\omega_\mathrm{in}$ [panel (b)] for a finite $\gamma$, the distribution is concentrated along a stripe parallel to the resonant diagonal line $\omega_{\rm in}=\omega_0$ with a lateral thickness again determined by the modulation strength and the finite lifetime.

In Fig.\,\ref{fig:comparison_sine_square} we present experimental data for both sinusoidal and square modulation for identical peak-peak modulation amplitude and modulation frequency, confirming the theoretical expectations: even though the sinusoidally modulated signal exhibits numerous sidebands above the input frequency, it lacks the second emission maximum present in the square-wave case.

\section{Additional theoretical considerations}
\label{app:theory}

In this final appendix we show analytical formulas for the amplitudes of the frequency comb based on the mapping on the Bloch wave dynamics. 

\subsection{Bloch wave dynamics with long-range hopping}
In this first Subsection we recall the problem of the Bloch wave dynamics in a tilted lattice with long-range hopping and we give an analytical expression for the propagator that is going to be used in the next Subsection.

We consider a model Hamiltonian with an extended long-range hopping on a tilted lattice, where each site $n$ has a local energy of $n \Delta\varepsilon$.
\begin{align}
\hat{H}_\mathrm{WS} & = 
 \hat{N} + \hat{U} \, ,  \label{eq:Hamilton-1}\\
\hat{N}  &= \Delta\varepsilon \sum_n n \ket{n}\bra{n} \, , \\
\hat{U}  &= - 
\sum_{m}
 \sum_n 
   \left( \frac{U_m}{4}   \ket{n}\bra{n+m} + \mbox{h.c.} \right) \,,
\end{align}
with $U_m \in \mathbb{R}$, $U_m=U_{-m}$ and $U_{m=0}=0$. The eigenstates of the quasi-momentum are
\begin{equation}
\ket{k}= \sum_n \frac{e^{i k n}}{\sqrt{2\pi}}
\ket{n} 
\, , \quad \
\left( \ket{n} = \frac{1}{\sqrt{2\pi}} 
\int^{\pi}_{-\pi} \!\!\! dk  e^{-i k n} \ket{k} \right) \, ,
\end{equation}
for $k \in \left[-\pi,\pi \right[$ in the first Brillouin zone.
These states are orthogonal $\bra{k}\ket{k'} = 2\pi 
\delta_{2\pi}\left( k-k'\right)$
with the periodic delta
$\delta_{2\pi}\left( x \right) = 
\sum_q \delta\left( x +2\pi q \right)$ ($q$ integer).
The operator has the following matrix element in the $k$ representation 
\begin{equation}
\bra{k}\hat{N}\ket{k'} 
=
i \frac{\partial }{\partial k} 
\delta_{2\pi}\left( k-k'\right) \, ,
\end{equation}
whereas the hopping operator in the $k$ representation is
\begin{equation}
\bra{k} \hat{U} \ket{k'}
 =
 U\left( k \right) 
 \delta_{2\pi}\left( k-k'\right)
 \,,
\end{equation}
with
\begin{equation}
U\left( k \right) 
=
\sum_{m > 0  } U_m \cos\left(m \, k  \right) 
 \, .
\end{equation}
The generic state on the lattice can be expressed using the position eigenstates $\ket{\Psi} = \sum_n c_n \ket{n}$ (with normalization $\sum_n {|c_n|}^2=  1$) or the quasi-momentum eigenstates $\ket{\Psi} = \int^{\pi}_{-\pi}\!\! dk  \phi\left(k\right) \ket{k}$ 
(with normalization 
$\int^{\pi}_{-\pi}\!\! dk {|  \phi\left(k\right)   |}^2=1$).
The coefficients $c_n$ are related to $ \phi\left(k\right) $ through the relations 
\begin{equation}
c_n =  \int^{\pi}_{-\pi}\!\! \frac{dk}{\sqrt{2\pi}}  e^{ikn} \phi\left(k\right)  \, , \quad
\phi\left(k\right) = \sum_n \frac{e^{-ikn}}{\sqrt{2\pi}} c_n 
\, .
\end{equation}
The Schr{\"o}dinger's equation for the coefficients $c_n$ is 
\begin{equation}
i\hbar \frac{\partial c_n}{\partial t }
=
\Delta\varepsilon \, n \, c_n - \sum_m \frac{U_m}{4}\left( c_{n+m} + c_{n-m} \right)
 \label{eq:Hwd_lattice}
\,,
\end{equation}
whereas the Schr{\"o}dinger's equation for 
$ \phi\left(k\right)$ is 
\begin{equation}
i\hbar \frac{\partial \phi\left(k\right)}{\partial t }  =  i \Delta\varepsilon \frac{\partial  \phi\left(k\right) }{\partial k} - U\left(k\right) \phi\left(k\right) \,.
\end{equation}
and the energy eigenstates are solution of the equation 
\begin{equation}
 i \Delta\varepsilon \frac{\partial  \phi\left(k\right) }{\partial k} - U\left(k\right) \phi\left(k\right)  = E  \phi\left(k\right)  
\end{equation}
corresponding to 
\begin{equation}
  \phi\left(k\right)  
 =  \phi\left(-\pi\right) e^{ -i\left(k+\pi \right)  \frac{E}{\Delta\varepsilon}  -i \int_{-\pi}^k \! dk' \frac{ U\left( k'  \right)}{\Delta\varepsilon} }
 \, .
 \end{equation}
As $k$ is restricted to the first Brillouin zone and the function $U\left( k  \right)$ is periodic $U\left( k  \right)=U\left( k +2\pi \right)$, the wavefunction is periodic in the $k$ representation 
$\phi\left(k\right)  
 =
 \phi\left(k+2\pi\right)  $
which implies that the spectrum is discrete
\begin{equation}
 E \longrightarrow E_{n_0} = n_o \Delta\varepsilon \, .
 \end{equation}
Fixing the normalization constant, up to a global phase, the eigenstates are 
\begin{equation}
 \phi_{n_0}\left(k\right)  
 = \frac{1}{\sqrt{2\pi}}
 e^{ -i k n_0  -i \int_{-\pi}^{k}  \! dk' \frac{ U\left( k'  \right)}{\Delta\varepsilon} }
 \, .
 \end{equation}
We now analyze the propagator defined as 
\begin{align}
& G_{n,m}\left( t  \right)
\!=\!\!
\bra{n} e^{-\frac{i}{\hbar} \hat{H}_\mathrm{WS} t} \ket{m}
=\!\!
\sum_{n_0} \langle n | \phi_{n_0}\rangle \langle{ \phi_{n_0}}| m\rangle  e^{-\frac{i}{\hbar} E_{n_0}   t}
\nonumber\\
&
\!
=
\!\!
\sum_{n_0} 
\!\!
 \iint^{\pi}_{-\pi}\!\!\! \frac{dk_1 dk_2}{4\pi^2}
  e^{i \left[ k_1 n - k_2 m 
  - \int_{k_2}^{k_1}  \! dk' \frac{ U\left( k'  \right)}{\Delta\varepsilon} \right]}
  \!
  e^{-i n_0\left(k_1-k_2 + \frac{\Delta\varepsilon}{\hbar}t \right)} 
  \nonumber\\
&= 
\!\!\!\!
\iint^{\pi}_{-\pi} \!\!\!\! \frac{dk_1 dk_2}{2\pi}
 e^{i \left[ k_1 n - k_2 m - \int_{k_2}^{k_1}  \! dk' \frac{ U\left( k'  \right)}{\Delta\varepsilon} \right]}
\delta_{2\pi}\!\left(\! k_2 \!- k_1\! - \frac{\Delta\varepsilon} {\hbar}   t \!\right) 
 .
\end{align}

The propagator $G_{n,m}\left( t  \right)$ is a periodic function with period $T=2\pi\hbar/\Delta\varepsilon$ corresponding to the period of the Bloch oscillations in the tilted lattice.

For $t=0$ or $t=T$, the function $\delta_{2\pi}\left( x \right)$ is equivalent to $\delta \left( x \right)$ as this is the component that produces a non-zero value within the double integral with the integration range defined by the square $k_1\in\left[ -\pi,\pi \right[$ and $k_2\in\left[ -\pi,\pi \right[$. This leads to the result $G_{n,m}(0)=\delta_{n,m}$.

For finite time, $0<t<T$, the function $\delta_{2\pi}\left( x \right)$ must be replaced with 
$\delta \left( x \right)$ and 
$\delta \left( x - 2\pi \right)$ 
since these two functions define two lines that are contained in the integration square range.
For example, we have
\begin{equation}
G_{n,0}\left( t  \right) 
\!
=
\!
\frac{1}{\pi}
\!
\int^{\pi}_{-\pi}\!\!\!\!\!\! dk_1
e^{i \left[ k_1 n + \int^{k_1+\frac{\Delta\varepsilon}{\hbar}t}_{k_1}  \! dk' \frac{ U\left( k'  \right)}{\Delta\varepsilon} \right]}
\,\, \left(0<t<T\right) \, .
\end{equation}
We now focus on the behavior of the propagator at half period $t=T/2$, which allows us to determine the width of the Bloch oscillations
\begin{equation}
G_{n,0}\left( \frac{T}{2}  \right) 
=
\frac{1}{\pi}
\int^{\pi}_{-\pi}\!\! dk_1
e^{i \left[ k_1 n + \int^{k_1+\pi}_{k_1}  \! dk' \frac{ U\left( k'  \right)}{\Delta\varepsilon} \right]}
\, .
\end{equation}
If we consider $U\left( k \right)$ as a piecewise step function with 
\begin{equation}
U\left( k \right)=   
\left\{
\begin{array}{l}
  -\frac{\Delta U_0}{2}    \qquad \mbox{for} \quad  \pi/2  < |k| <  \pi
\\
  \frac{\Delta U_0}{2}     \qquad  
\mbox{for} \quad  | k|  < \pi/2
\end{array}
\right.
\, ,
\label{eq:U_k}
\end{equation}
the integral of the function $U\left( k \right)$ is a piecewise of linear functions
that leads to the following result for the propagator
\begin{align}
& G_{n,0}\left(\frac{T}{2} \right) 
= \nonumber \\
&
= \frac{2}{\pi} 
\left\{
\frac{
\sin\left[\frac{\pi}{2} \left( n - \frac{\Delta U_0}{\Delta\varepsilon}  \right) \right]
}
{ n-\frac{\Delta U_0}{\Delta\varepsilon} }
+
{\left(-1 \right)}^{n}
\frac{
\sin\left[ \frac{\pi}{2} \left( n + \frac{\Delta U_0}{\Delta\varepsilon}  \right) \right]
}
{ n+ \frac{\Delta U_0}{\Delta\varepsilon} }
\right\}
\, .
\end{align}
showing that the propagator is peaked at $\pm \Delta U_0/\Delta \varepsilon$.

\subsection{Mapping of the modulated resonator to the Bloch wave dynamics}

In this final Subsection we make use of the propagator to provide a formal solution for the amplitude of the frequency comb in the experimentally relevant case.

We consider the motion equation for the field amplitude
$\alpha(t)$ of the resonator with $\delta\omega(t)=\delta\omega_\mathrm{pp}\,\Theta\left[\cos\left(\Omega t\right)\right]$.
Setting $\delta\tilde{\omega}(t)= \delta\omega(t)-\delta\omega_\mathrm{pp}/2$ we write
\begin{equation}
  i \dot{\alpha} \! \left( t \right)
 =
 \left[\omega_0+
  \frac{\delta\omega_\mathrm{pp}}{2}
  + \delta\tilde{\omega}(t)- i \frac{ \gamma }{ 2 }\right] \alpha \! \left( t \right)
  + E_\mathrm{in} e^{-i\omega_{\mathrm{in}} t} \, .
\end{equation}
Inserting the expression $\alpha(t) = \sum_n \tilde{\alpha}_n(t)
e^{-i \left( \omega_{\mathrm{in}} +  n \Omega  \right) t}$, the equation for the coefficients $\tilde{\alpha}_n(t)$ is
\begin{align}
&  
i \dot{\tilde{\alpha}}_n \! \left( t \right) 
= 
-  n \Omega  \tilde{\alpha}_n \! \left( t \right)
\nonumber \\
&
+
\sum_m \delta\tilde{\omega}_m \tilde{\alpha}_{n-m} (t)
+
\left[
  \frac{\delta\omega_\mathrm{pp}}{2} - \Delta 
- i \frac{ \gamma }{ 2 }\right]
  \tilde{\alpha}_n \! \left( t \right)
  + 
E_\mathrm{in} \delta_{n,0}
 \, ,
 \label{eq:beta_n}
\end{align}
with $\Delta = \omega_\mathrm{in} - \omega_0 $ 
and $\delta\tilde{\omega}(t) = \sum_m \delta\tilde{\omega}_m e^{-i n \Omega t}$.
The stationary coefficients $\{\alpha_n\}$ are solutions of the equation $\dot{\tilde{\alpha}}_n(t)=0$.

We can write Eq.\,(\ref{eq:beta_n}) using a vector/matrix representation as
\begin{equation}
\label{eq:nonSCHRO}
i \frac{\partial }{\partial t}
\ket{\tilde{\alpha}} = 
\tilde{H}_\mathrm{WS} \ket{\tilde{\alpha}}
+
\left[
  \frac{\delta\omega_\mathrm{pp}}{2} - \Delta 
- i \frac{ \gamma }{ 2 }\right] 
\ket{\tilde{\alpha}}
+  \ket{E_\mathrm{in}}
\end{equation}
with $\tilde{\alpha}=\langle n | \tilde{\alpha} \rangle$, 
$ 
\langle n |E_\mathrm{in} \rangle  = E_\mathrm{in} \delta_{n,0}
$
and $\tilde{H}_\mathrm{WS}$ is the Hamiltonian 
Eq.\,(\ref{eq:Hamilton-1}) with the substitution
\begin{equation}
    \frac{\Delta\varepsilon}{\hbar} \rightarrow  -\Omega \,, \qquad \frac{U_m}{\hbar} \rightarrow -2\delta\tilde{\omega}_m \, .
    \label{eq:Hwd-mapped}
\end{equation}
The formal solution of Eq.\,(\ref{eq:Hwd-mapped}) reads
\begin{equation}
    \ket{\tilde{\alpha}} =
    \int^t_{t_0}\!\!\! dt' 
    e^{- i\left( \tilde{H}_\mathrm{WS} +
    \left[
  \frac{\delta\omega_\mathrm{pp}}{2} - \Delta 
- i \frac{ \gamma }{ 2 }\right] 
    \right) (t-t') }
    \ket{E_\mathrm{in} } \, .
\end{equation}
We let the time $t_0\rightarrow -\infty$ and use the expansion in terms of the eigenstates on the lattice of the matrix/Hamiltonian $\tilde{H}_\mathrm{WS}$ to obtain the stationary solution
\begin{align}
&\alpha_n  = \langle n | \alpha \rangle= \nonumber \\
&
=
  \int^t_{-\infty}\!\!\! dt' 
  e^{ 
    \left[-i
  \frac{\delta\omega_\mathrm{pp}}{2} - \Delta 
- i \frac{ \gamma }{ 2 }\right] 
(t-t') }
\langle n |
e^{- i \tilde{H}_\mathrm{WS}  (t-t') }
    \ket{E_\mathrm{in} }
    \nonumber \\
&
\!\!\! 
=
\!\!  \int^t_{-\infty}\!\!\!\!\!\!\!\! dt' 
  e^{ -i 
    \left[
  \frac{\delta\omega_\mathrm{pp}}{2} - \Delta 
- i \frac{ \gamma }{ 2 }\right] 
(t-t') }
\sum_{n_0} e^{i n_0 \Omega   \left(t-t' \right)}
\langle n |  \tilde{\phi}_{n_0} \rangle
\langle \tilde{\phi}_{n_0}  | 0 \rangle 
E_\mathrm{in}  
\nonumber\\
&
=
E_\mathrm{in}
  \int^t_{-\infty}\!\!\! dt' 
  e^{ -i 
    \left[
  \frac{\delta\omega_\mathrm{pp}}{2} - \Delta 
- i \frac{ \gamma }{ 2 }\right] 
(t-t') }
G_{n,0}\left( t-t'\right)
\nonumber\\
&
=
E_\mathrm{in}
  \int^{\infty}_{0}\!\!\! d\tau
  e^{ -i 
    \left[
  \frac{\delta\omega_\mathrm{pp}}{2} - \Delta 
- i \frac{ \gamma }{ 2 }\right] 
\tau }
G_{n,0}\left( \tau \right) \, .
\label{eq:sol}
\end{align}
To capture the qualitative behavior of the frequency comb, we consider the limit $\delta\omega_\mathrm{pp}/2 - \Delta =0 $ 
for which we have 
\begin{equation}
\tilde{\alpha}_n  
=
E_\mathrm{in}
  \int^{\infty}_{0}\!\!\! d\tau 
  e^{ -  \frac{ \gamma }{ 2 }  \tau }
G_{n,0}\left(\tau\right)
\label{eq:delta=0}
\end{equation}
From the previous Subsection, we have seen that
$G_{n,0}$ is peaked at distance $\pm \delta\omega_\mathrm{pp}/2 \Omega$ with respect to the starting point for times close to the half period.
These terms represent the main contribution to the time integration in Eq.\,(\ref{eq:delta=0}).

At arbitrary detuning we must consider the general solution Eq.\,(\ref{eq:sol}) with 
$\delta\omega_\mathrm{pp}/2 - \Delta  \neq 0 $.
We can always write 
\begin{equation}
\Delta - \frac{\delta\omega_\mathrm{pp}}{2}  
= \bar{n} \Omega + \delta\varphi
\end{equation}
with $ \delta\varphi \ll 1$, and the equation for $\tilde{\alpha}(t)$ becomes 
\begin{equation}
  i \dot{\tilde{\alpha}} \! \left( t \right)
  = 
 \left[
   - \bar{n} \Omega - \delta\varphi 
  + \delta\tilde{\omega}(t)- i \frac{ \gamma }{ 2 }\right]
  \tilde{\alpha} \! \left( t \right)
  + E_\mathrm{in}  
\, .
\end{equation}
Then we apply a further transformation 
$\tilde{\alpha} = e^{i \bar{n} \Omega t } \beta(t)$ with 
the coefficients $\beta$ satisfying the following equations 
\begin{equation}
  i \dot{\beta} \! \left( t \right)
  = 
 \left[ 
  \delta\tilde{\omega}(t)- i \frac{ \gamma }{ 2 }\right]
  \beta \! \left( t \right)
  + E_\mathrm{in}  e^{-i  \bar{n} \Omega t } 
\, .
\end{equation}
\newline\noindent Using the ansatz $\beta=\sum_n \beta_n e^{-i\Omega n t}$, we have 
\begin{equation}
i \dot{\beta}_n \! \left( t \right) 
=
\sum_m \delta\tilde{\omega}_m \beta_{n-m} \left( t \right)
+
\left( -\delta \varphi - i \frac{ \gamma }{ 2 } \right)  \beta_{n} \left( t \right)
+ 
E_\mathrm{in} \delta_{n,\bar{n}} 
\, .
\end{equation}
Repeating the same steps, one obtains the stationary solution
\begin{align}
\beta_n   
=
E_\mathrm{in}
  \int^{\infty}_{0}\!\!\! d\tau
  e^{ -i 
    \left[ \delta\varphi 
- i \frac{ \gamma }{ 2 }\right] 
\tau }
G_{n,\bar{n}}\left( \tau \right) \, .
\end{align}
In other words, the effect of finite detuning is equivalent to a shift of the source term, which implies a frequency shift of the solutions in the frequency combs controlled by the tuning, as shown in the main text.

\bibliography{bliblio}

\end{document}